\documentclass[prodmode,acmcacm]{acmsmall} 

\usepackage{epstopdf}
\usepackage{soul}

\usepackage[ruled]{algorithm2e}

\SetAlFnt{\small}
\SetAlCapFnt{\small}
\SetAlCapNameFnt{\small}
\SetAlCapHSkip{0pt}
\IncMargin{-\parindent}

\acmVolume{X}
\acmNumber{X}
\acmArticle{XX}
\acmYear{201X}
\acmMonth{0}

\begin{document}

\markboth{E. Ferrara et al.}{The Rise of Social Bots}

\title{The Rise of Social Bots}
\author{EMILIO FERRARA
\affil{Indiana University}
ONUR VAROL
\affil{Indiana University}
CLAYTON DAVIS
\affil{Indiana University}
FILIPPO MENCZER
\affil{Indiana University}
ALESSANDRO FLAMMINI
\affil{Indiana University}
}

\begin{abstract}
The Turing test aimed to recognize the behavior of a human from that of a computer algorithm.
Such challenge is more relevant than ever in today's social media context, where limited attention and technology constrain the expressive power of humans, while incentives abound to develop software agents mimicking humans.  These social bots interact, often unnoticed, with real people in  social media ecosystems, but their abundance is uncertain.
While many bots are benign, one can design harmful bots with the goals of persuading, smearing, or deceiving. 
Here we discuss the characteristics of modern, sophisticated social bots, and how their presence can endanger online ecosystems and our society.
We then review current efforts to detect social bots on Twitter. Features related to content, network, sentiment, and temporal patterns of activity are imitated by bots but at the same time can help discriminate synthetic behaviors from human ones, yielding signatures of engineered social tampering. 
\end{abstract}

\category{}{Human-centered computing}{Collaborative and social computing}[Social media]
\category{}{Information systems}{World Wide Web}[Social networks]
\category{}{Networks}{Network types}[Social media networks]


\acmformat{Emilio Ferrara, Onur Varol, Clayton Davis, Filippo Menczer, and Alessandro Flammini. 2015. The Rise of Social Bots.}

\begin{bottomstuff}
This work is supported in part by the National Science Foundation (grant CCF-1101743), DARPA (grant W911NF-12-1-0037), and the James McDonnell Foundation (grant 220020274). The funders had no role in study design, data collection and analysis, decision to publish, or preparation of the manuscript.

Corresponding author: E. Ferrara (ferrarae@indiana.edu)

Author's address: Center for Complex Networks and Systems Research. School of Informatics and Computing. Indiana University Bloomington. 919 E. 10th Street, Bloomington, IN 47408, USA
\end{bottomstuff}

\maketitle

\section*{The rise of the machines}

Bots (short for \emph{software robots}) have been around since the early days of computers: one compelling example is that of \emph{chatbots},  algorithms designed to hold a conversation with a human, as envisioned by Alan Turing in the 1950s~\cite{turing1950computing}.
The dream of designing a computer algorithm that passes the Turing test has driven artificial intelligence research for decades, as witnessed by initiatives like the Loebner Prize, awarding progress in natural language processing.\footnote{\url{www.loebner.net/Prizef/loebner-prize.html}}  
Many things have changed since the early days of AI, when bots like Joseph Weizenbaum's ELIZA~\cite{weizenbaum1966eliza}, mimicking a Rogerian psychotherapist, were developed as demonstrations or for delight. 

Today, social media ecosystems populated by hundreds of millions of individuals present real incentives ---including economic and political ones--- to design algorithms that exhibit human-like behavior. Such ecosystems also raise the bar of the challenge, as they introduce new dimensions to emulate in addition to content, including the social network, temporal activity, diffusion patterns and sentiment expression. 
A \emph{social bot} is a computer algorithm that automatically produces content and interacts with humans on social media, trying to emulate and possibly alter their behavior. Social bots have been known to inhabit social media platforms for a few years~\cite{lee2011seven,boshmaf2013design}.

\section*{Engineered social tampering}

What are the intentions of social bots? Some of them are benign and, in principle, innocuous or even helpful: this category includes bots that automatically aggregate content from various sources, like simple news feeds. Automatic responders to inquiries are increasingly adopted by brands and companies for customer care. 
Although this type of bots are designed to provide a useful service, they can sometimes be harmful, for example when they contribute to the spread of unverified information or rumors. 
Analysis of Twitter posts around the Boston marathon bombing revealed that social media can play an important role in the early recognition and characterization of emergency events~\cite{cassa2013twitter}. But false accusations also circulated widely on Twitter in the aftermath of the attack, mostly due to bots automatically retweeting posts without verifying the facts or checking the credibility of the source~\cite{gupta20131}.

With every new technology comes abuse, and social media are no exception. 
A second category of social bots includes malicious entities designed specifically with the purpose to harm. 
These bots mislead, exploit, and manipulate social media discourse with rumors, spam, malware, misinformation, slander, or even just noise. 
This may result in several levels of damage to society. For example, bots may artificially inflate support for a political candidate~\cite{ratkiewicz2011detecting}; such activity could endanger democracy by influencing the outcome of elections.
In fact, these kinds of abuse have already been observed: during the 2010 U.S. midterm elections, social bots were employed to support some candidates and smear their opponents, injecting thousands of tweets pointing to websites with fake news~\cite{ratkiewicz2011detecting}. A similar case was reported around the Massachusetts special election of 2010~\cite{metaxas2012social}. Campaigns of this type are sometimes referred to as astroturf or Twitter bombs.  
The problem is not just in establishing the veracity of the information being promoted --- this was an issue before the rise of social bots, and remains beyond the reach of algorithmic approaches. The novel challenge brought by bots is the fact that they can give the false impression that some piece of information, regardless of its accuracy, is highly popular and endorsed by many, exerting an influence against which we haven't yet developed antibodies. Our vulnerability makes it possible for a bot to acquire significant influence, even unintentionally~\cite{aiello2012people}.
Sophisticated bots can generate personas that appear as credible followers, and thus are harder for both people and filtering algorithms to detect. They make for valuable entities on the fake follower market, and allegations of acquisition of fake followers have touched several prominent political figures in the US and worldwide. 

More examples of the potential dangers brought by social bots are increasingly reported by journalists, analysts, and researchers. These include the unwarranted consequences that the widespread diffusion of bots may have on the stability of markets.
There have been claims that Twitter signals can be leveraged to predict the stock market~\cite{bollen2011twitter}, and there is  an increasing amount of evidence showing that market operators pay attention and react promptly to information from social media. On April 23, 2013, for example, the Syrian Electronic Army hacked the Twitter account of the Associate Press and posted a false rumor about a terror attack on the White House in which President Obama was allegedly injured. This provoked an immediate crash in the stock market. 
On May 6, 2010 a \emph{flash crash} occurred in the U.S. stock market, when the Dow Jones plunged over 1,000 points (about 9\%) within minutes ---the biggest one-day point decline in history. After a 5-month long investigation, the role of high-frequency trading bots became obvious, but it yet remains unclear whether these bots had access to information from the social Web~\cite{hwang2012socialbots}.
The combination of social bots with an increasing reliance on automatic trading systems that, at least partially, exploit information from social media, is ripe with risks. 
Bots can amplify the visibility of misleading information, while automatic trading system lack  fact-checking capabilities.
A recent orchestrated bot campaign successfully created the appearance of a sustained discussion about a tech company called Cynk. Automatic trading algorithms picked up this conversation and started trading heavily in the company's stocks. This resulted in a 200-fold increase in market value, bringing the company's worth to 5 billion dollars.\footnote{The Curious Case of Cynk, an Abandoned Tech Company Now Worth \$5 Billion --- \url{mashable.com/2014/07/10/cynk}} By the time analysts recognized the orchestration behind this operation and stock trading was suspended, the losses were real.

\section*{The bot effect}

These anecdotes illustrate the consequences that tampering with the social Web may have for our increasingly interconnected society.
In addition to potentially endangering democracy, causing panic during emergencies, and affecting the stock market, social bots can harm our society in even subtler ways. 
A recent study demonstrated the vulnerability of social media users to a \emph{social botnet} designed to expose private information, like phone numbers and addresses~\cite{boshmaf2013design}. 
This kind of vulnerability can be exploited by cybercrime and cause the erosion of trust in social media~\cite{hwang2012socialbots}.
Bots can also hinder the advancement of public policy by creating the impression of a grassroots movement of contrarians, or contribute to the strong polarization of political discussion observed in social media~\cite{conover2011political}.
They can alter the perception of social media influence, artificially enlarging the audience of some people~\cite{edwards2014bot}, or they can ruin the reputation of a company, for commercial or political purposes~\cite{messias2013you}.
A recent study demonstrated that emotions are contagious on social media~\cite{kramer2014experimental}: elusive bots could easily infiltrate a population of unaware humans and manipulate them to affect their perception of reality, with unpredictable results.
Indirect social and economic effects of social bot activity include the alteration of social media analytics, adopted for various purposes such as TV ratings,\footnote{Nielsen's New Twitter TV Ratings Are a Total Scam. Here's Why. --- \url{defamer.gawker.com/nielsens-new-twitter-tv-ratings-are-a-total-scam-here-1442214842}} expert finding~\cite{wu2013detecting}, and scientific impact measurement.\footnote{altmetrics: a manifesto --- \url{altmetrics.org/manifesto/}}

\section*{Act like a human, think like a bot}

One of the greatest challenges for bot detection in social media is in understanding what modern social bots can do~\cite{boshmaf2012key}. 
Early bots mainly performed one type of activity: posting content automatically. 
These bots were as naive as easy to spot by trivial detection strategies, such as focusing on high volume of content generation. 
In 2011, James Caverlee's team at Texas A\&M University implemented a honeypot trap that managed to detect thousands of social bots~\cite{lee2011seven}. 
The idea was simple and effective: the team created a few Twitter accounts (bots) whose role was solely to create nonsensical tweets with gibberish content, in which no human would ever be interested. 
However, these accounts attracted many followers. 
Further inspection confirmed that the suspicious followers were indeed social bots trying to grow their social circles by blindly following random accounts. 

In recent years, Twitter bots have become increasingly sophisticated, making their detection more difficult. 
The boundary between human-like and bot-like behavior is now fuzzier. 
For example, social bots can search the Web for information and media to fill their profiles, and post collected material at predetermined times, emulating the human temporal signature of content production and consumption ---including circadian patterns of daily activity and temporal spikes of information generation~\cite{golder2011diurnal}. 
They can even engage in more complex types of interactions, such as entertaining conversations with other people, commenting on their posts, and answering their questions~\cite{hwang2012socialbots}. 
Some bots specifically aim to achieve greater influence by gathering new followers and expanding their social circles; they can search the social network for popular and influential people and follow them or capture their attention by sending them inquiries, in the hope to be noticed~\cite{aiello2012people}.
To acquire visibility, they can infiltrate popular discussions, generating topically-appropriate ---and even potentially interesting--- content, by identifying relevant keywords and searching online for information fitting that conversation~\cite{freitas2014reverse}. 
After the appropriate content is identified, the bots can automatically produce responses through natural language algorithms, possibly including references to media or links pointing to external resources. 
Other bots aim at tampering with the identities of legitimate people: some are  identity thieves, adopting slight variants of real usernames, and stealing personal information such as pictures and links.
Even more advanced mechanisms can be employed; some social bots are able to ``clone'' the behavior of legitimate users, by interacting with their friends and posting topically coherent content with similar temporal patterns.

\section*{A taxonomy of social bot detection systems}
For all the reasons outlined above, the computing community is engaging in the design of advanced methods to automatically detect social bots, or to discriminate between humans and bots. 
The strategies currently employed by social media services appear inadequate to contrast this phenomenon and the efforts of the academic community in this direction just started.

In the following, we propose a simple taxonomy that divides the approaches proposed in literature into three classes: \emph{(i)} bot detection systems based on social network information; \emph{(ii)} system based on crowd-sourcing and leveraging human intelligence; \emph{(iii)} machine learning methods based on the identification of highly-revealing features that discriminate between bots and humans. Sometimes a hard categorization of a detection strategy into one of these three categories is difficult, since some exhibit mixed elements: we present also a section of methods that combine ideas from these three main approaches.

\subsection*{Graph-based social bot detection}

The challenge of social bot detection has been framed by various teams in an adversarial setting~\cite{alvisi2013sok}. One example of this framework is represented by the Facebook Immune System~\cite{stein2011facebook}: an adversary may control multiple social bots (often referred to as \emph{Sybils} in this context) to impersonate different identities and launch an attack or infiltration. Proposed strategies to detect \emph{sybil accounts} often rely on examining the structure of a social graph. SybilRank~\cite{cao2012aiding} for example assumes that sybil accounts exhibit a small number of links to legitimate users, instead connecting mostly to other sybils, as they need a large number of social ties to appear trustworthy. This feature is exploited to identify densely interconnected groups of sybils. 
One common strategy is to adopt off-the-shelf community detection methods to reveal such tightly-knit local communities; however, the choice of the community detection algorithm has proven to crucially affect the performance of the detection algorithms~\cite{viswanath2011analysis}.
A wise attacker may counterfeit the connectivity of the controlled sybil accounts so that to mimic the  features of the community structure of the portion of the social network populated by legitimate accounts; this strategy would make the attack invisible to methods solely relying on community detection. 
To address this shortcoming, some detection systems, for example SybilRank, also employ the paradigm of \emph{innocent by association}: an account interacting with a legitimate user is considered itself legitimate. Souche~\cite{xie2012innocent} and Anti-Reconnaissance~\cite{paradise2014anti} also rely on the assumption that social network structure alone separates legitimate users from bots. 
Unfortunately, the effectiveness of such detection strategies is bound by the behavioral assumption that legitimate users refuse to interact with unknown accounts. This was proven unrealistic by various experiments~\cite{stringhini2010detecting,boshmaf2013design,elyashar2013homing}: a large-scale social bot infiltration on Facebook showed that over 20\% of legitimate users accept friendship requests indiscriminately, and over 60\% accept requests from accounts with at least one contact in common~\cite{boshmaf2013design}. On other platforms like Twitter and Tumblr, connecting and interacting with strangers is one of the main features. In these circumstances, the innocent-by-association paradigm yields high false positive rates, and these are the worst types of errors: a service provider would rather fail to detect a social bot than inconvenience a real user with an erroneous account suspension.
Some authors noted the limits of the assumption of finding groups of social bots or legitimate users only: real platforms may contain many mixed groups of legitimate users who fell prey of some bots~\cite{alvisi2013sok}, and sophisticated bots may succeed in large-scale infiltrations making it impossible to detect them solely from network structure information. This brought Alvisi et al. to recommend a portfolio of complementary detection techniques, and the manual identification of legitimate social network users to aid in the training of supervised learning algorithms.

\subsection*{Crowd-sourcing social bot detection}
The possibility of human detection has been explored by Wang et al.~\cite{wang2012social} who suggest the crowd-sourcing of social bot detection to legions of  workers. As a proof-of-concept, they created an Online Social Turing Test platform. The authors assumed that bot detection is a simple task for humans, whose ability to evaluate conversational nuances like sarcasm or persuasive language, or to observe emerging patterns and anomalies, is yet unparalleled by machines. Using data from Facebook and Renren (a popular Chinese online social network), the authors tested the efficacy of humans, both expert annotators and workers hired online, at detecting social bot accounts simply from the information on their profiles. 
The authors observed that the detection rate for hired workers drops off over time, although it remains good enough to be used in a majority voting protocol: the same profile is shown to multiple workers and the opinion of the majority determines the final verdict. This strategy exhibits a near-zero false positive rate, a very desirable feature for a service provider. 
Three drawbacks undermine the feasibility of this approach: first, although the authors make a general claim that crowd-sourcing the detection of social bots might work if implemented since the early stage, this solution might not be cost-effective for a platform with a large pre-existing user base, like Facebook and Twitter. 
Second, to guarantee that a minimal number of human annotators can be employed to minimize costs, ``expert'' workers are still needed to accurately detect fake accounts, as the ``average'' worker does not perform well individually. As a result, to reliably build a ground-truth of annotated bots, large social network companies like Facebook and Twitter are forced to hire teams of expert analysts~\cite{stein2011facebook}, however such a choice might not be suitable for small social networks in their early stages (an issue at odds with the previous point).
Finally, exposing personal information to external workers for validation raises privacy issue~\cite{elovici2013ethical}. While Twitter profiles tend to be more public compared to Facebook, Twitter profiles also contain less information than Facebook or Renren, thus giving a human annotator less ground to make a judgment. 
Analysis by manual annotators of interactions and content produced by a Syrian social botnet active in Twitter for 35 weeks suggests that some advanced social bots may no longer aim at mimicking human behavior, but rather at misdirecting attention to irrelevant information~\cite{abokhodair2014dissecting}. Such \emph{smoke screening} strategies require high coordination among the bots. This observation is in line with early findings on political campaigns orchestrated by social bots, which exhibited not only peculiar network connectivity patterns but also enhanced levels of coordinated behavior~\cite{ratkiewicz2011detecting}.
The idea of leveraging information about the synchronization of account activities has been fueling many social bot detection systems: frameworks like CopyCatch~\cite{beutel2013copycatch}, SynchroTrap~\cite{cao2014uncovering}, and the Renren Sybil detector~\cite{wang2013you,yang2014uncovering} rely explicitly on the identification of such coordinated behavior to identify social bots.

\begin{table*}
\centering
\tbl{Classes of features employed by feature-based  systems for social bot detection.}{
\begin{tabular}{@{}lp{5.1in}@{}}
Class &	Description \\
\hline\hline
Network &
\emph{Network features} capture various dimensions of information diffusion patterns. Statistical features can be extracted from networks based on \emph{retweets}, \emph{mentions}, and \emph{hashtag co-occurrence}. Examples include degree distribution, clustering coefficient, and centrality measures~\cite{ratkiewicz2011truthy}. 
\\
User &
\emph{User features} are based on Twitter meta-data related to an account, including language, geographic locations, and account creation time. 
\\
Friends &
\emph{Friend features} include descriptive statistics relative to an account's social contacts, such as median, moments, and entropy of the distributions of their numbers of followers, followees, and posts. 
\\
Timing &	
\emph{Timing features} capture temporal patterns of content generation (tweets) and consumption (retweets); examples include the signal similarity to a Poisson process~\cite{ghosh2011entropy}, or the average time between two consecutive posts. 
\\
Content & 	
\emph{Content features} are based on linguistic cues computed through natural language processing, especially part-of-speech tagging; examples include the frequency of verbs, nouns, and adverbs in tweets. 
\\
Sentiment &	
\emph{Sentiment features} are built using general-purpose and Twitter-specific sentiment analysis algorithms, including happiness, arousal-dominance-valence, and emotion scores~\cite{golder2011diurnal,bollen2011twitter}.
\\
\end{tabular}}
\label{tab:features}
\end{table*}

\subsection*{Feature-based social bot detection}

The advantage of focusing on behavioral patterns is that these can be easily encoded in features and adopted with machine learning techniques to learn the signature of human-like and bot-like behaviors. This allows to later classify accounts according to their observed behaviors. Different classes of features are commonly employed to capture orthogonal dimensions of users' behaviors, as summarized in Table~\ref{tab:features}.

\begin{figure}\centering
\includegraphics[width=\columnwidth]{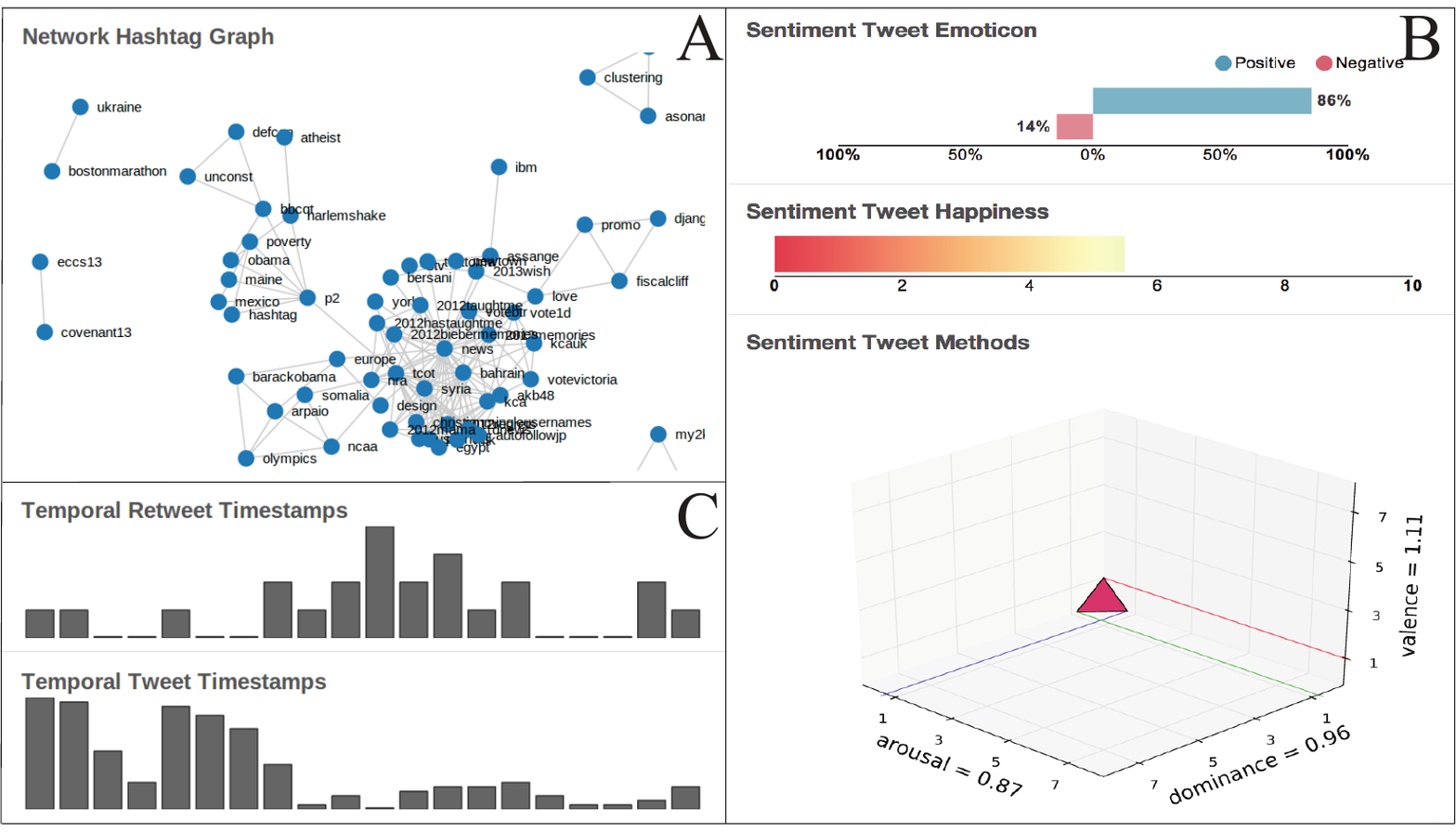}
\caption{Common features used for social bot detection. (A) The network of hashtags co-occurring in the tweets of a given user. (B) Various sentiment signals including emoticon, happiness and arousal-dominance-valence scores. (C) The volume of content produced and consumed (tweeting and retweeting) over time.}
\label{fig:panel}
\end{figure}

One example of feature-based system is represented by \emph{Bot or Not?}. Released in 2014, it was the  first social bot detection interface for Twitter to be made publicly available to raise  awareness about the presence of social bots.\footnote{As of the time of this writing, \emph{Bot or Not?} remains the only social bot detection system with a public-facing interface: \url{http://truthy.indiana.edu/botornot}}
Similarly to other feature-based systems~\cite{ratkiewicz2011truthy}, \emph{Bot or Not?} implements a detection algorithm relying upon highly-predictive features which capture a variety of suspicious behaviors and well separate social bots from humans. 
The system employs off-the-shelf supervised learning algorithms trained with examples of both humans and bots behaviors, 
based on the Texas A\&M dataset~\cite{lee2011seven} 
that contains 15 thousand examples of each class and millions of tweets. 
\emph{Bot or Not?} scores a detection accuracy above 95\%,\footnote{Detecting more recent and sophisticated social bots, compared to those in the 2011 dataset, may well yield lower accuracy.} measured by AUROC via cross validation.
In addition to the classification results, \emph{Bot or Not?} provides a variety of interactive visualizations that provide insights on the features exploited by the system (see Fig.~\ref{fig:panel} for examples).

\begin{figure}\centering
\includegraphics[width=0.65\columnwidth]{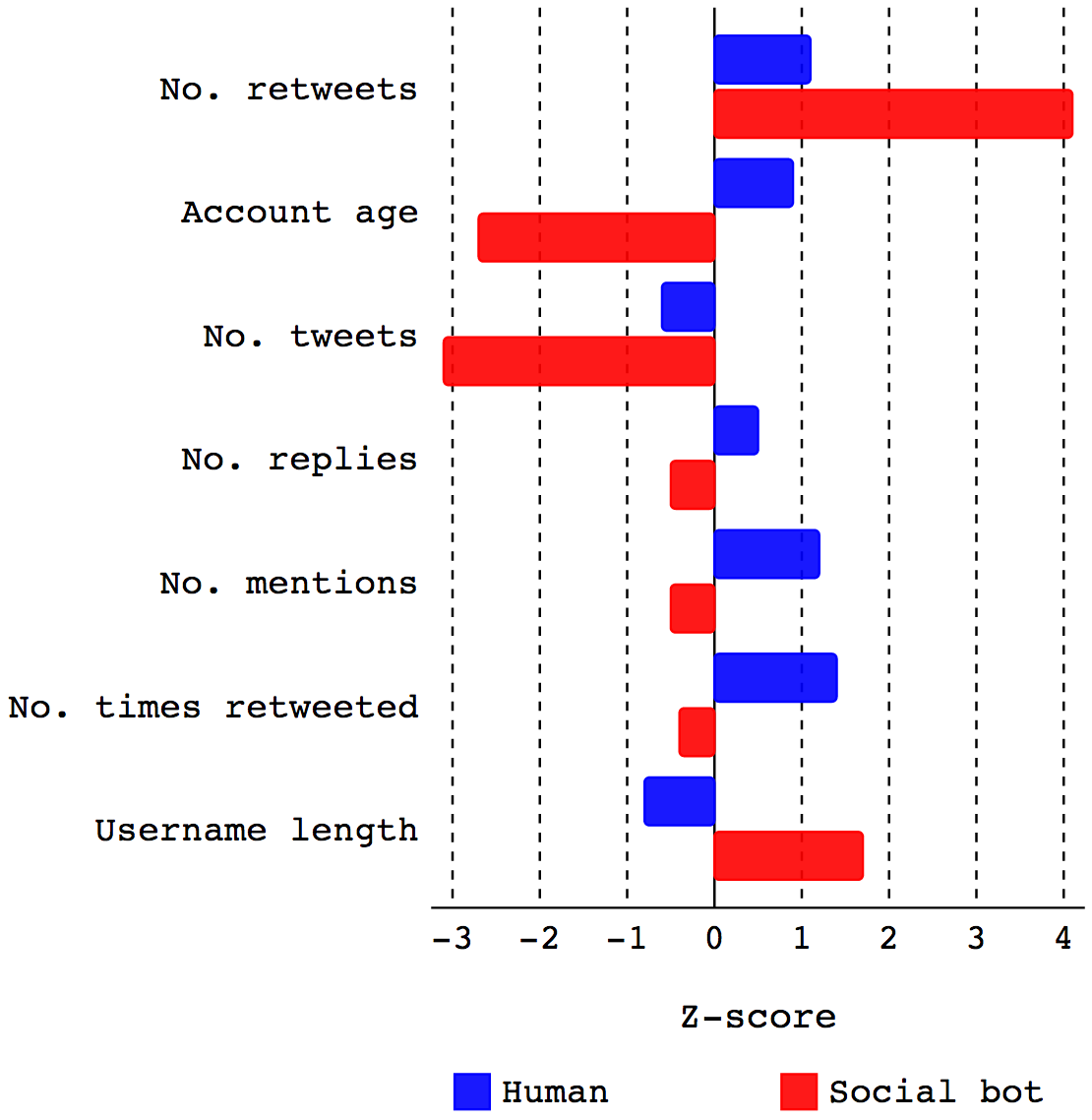}
\caption{User behaviors that best discriminate social bots from humans. Social bots retweet more than humans and have longer user names, while they produce fewer tweets, replies and mentions, and they are retweeted less than humans. Bot accounts also tend to be more recent.}
\label{fig:zscores}
\end{figure}

Bots are continuously changing and evolving: 
the analysis of the highly-predictive behaviors that feature-based detection systems can detect may reveal interesting patterns and provide unique opportunities to understand how to discriminate between bots and humans.
User meta-data are considered among the most predictive features and the most interpretable ones~\cite{hwang2012socialbots,wang2012social}: we can suggest few rules of thumb to infer whether an account is likely a bot, by comparing its meta-data with that of legitimate users (see Fig.~\ref{fig:zscores}).
Further work, however, will be needed to detect sophisticated strategies exhibiting a mixture of humans and social bots features (sometimes referred to as \emph{cyborgs}). Detecting these  bots, or hacked accounts~\cite{zangerle2014hacked}, is currently impossible for feature-based systems.

\subsection*{Combining multiple approaches}
Alvisi et al.~\cite{alvisi2013sok} recognized first the need of adopting complementary detection techniques to effectively deal with sybil attacks in social networks. The Renren Sybil detector~\cite{wang2013you,yang2014uncovering} is an example of system that explores multiple dimensions of users' behaviors like activity and timing information. Examination of ground-truth clickstream data shows that real users spend comparatively more time messaging and looking at other users' contents (such as photos and videos), whereas Sybil accounts spend their time harvesting profiles and befriending other accounts. Intuitively, social bot activities tend to be simpler in terms of variety of behavior exhibited.
By also identifying highly-predictive features such as invitation frequency, outgoing requests accepted, and network clustering coefficient, Renren is able to classify accounts into two categories: bot-like and human-like prototypical profiles~\cite{yang2014uncovering}. 
Sybil accounts on Renren tend to collude and work together to spread similar content: this additional signal, encoded as content and temporal similarity, is used to detect colluding accounts.
In some ways, the Renren approach~\cite{wang2013you,yang2014uncovering} combines the best of network- and behavior-based conceptualizations of Sybil detection. 
By achieving good results even utilizing only the last 100 click events for each user, the Renren system obviates to the need to store and analyze the entire click history for every user. Once the parameters are tweaked against ground truth, the algorithm can be seeded with a fixed number of known legitimate accounts and then used for mostly unsupervised classification. The ``Sybil until proven otherwise'' approach (the opposite of the innocent-by-association strategy) baked into this framework does lend itself to detecting previously unknown methods of attack: the authors recount the case of spambots embedding text in images to evade detection by content analysis and URL blacklists. Other systems implementing mixed methods, like CopyCatch~\cite{beutel2013copycatch} and SynchroTrap~\cite{cao2014uncovering}, also score comparatively low false positive rates with respect  to, for example, network-based methods.

\section*{Master of puppets}

If social bots are the puppets, additional efforts will have to be directed at finding  their ``masters.'' 
Governments\footnote{Russian Twitter political protests `swamped by spam' --- \url{www.bbc.com/news/technology-16108876}} and other entities with sufficient resources\footnote{Fake Twitter accounts used to promote tar sands pipeline --- \url{www.theguardian.com/environment/2011/aug/05/fake-twitter-tar-sands-pipeline}} have been alleged to use social bots to their advantage. Assuming the availability of effective detection technologies, it will be crucial to reverse-engineer the observed social bot strategies: who they target, how they generate content, when they take action, and what topics they talk about. 
A systematic extrapolation of such information may enable identification of the puppet masters.

Efforts in the direction of studying platforms vulnerability already started. Some researchers~\cite{freitas2014reverse}, for example, reverse-engineer social bots reporting alarming results: simple automated mechanisms that produce contents and boost  followers yield successful infiltration strategies and increase the social influence of the bots.
Others teams are creating bots themselves: Tim Hwang's ~\cite{hwang2012socialbots} and Sune Lehmann's\footnote{You are here because of a robot --- \url{sunelehmann.com/2013/12/04/youre-here-because-of-a-robot/}} groups continuously challenge our understanding of what strategies effective bots employ, and help quantify the susceptibility of people to their influence~\cite{wagner2012social,wald2013predicting}. Briscoe et al.~\cite{briscoe2014cues} studied the deceptive cues of language employed by influence bots. 
Tools like \emph{Bot or Not?} have been made available to the public to shed light on the presence of social bots online.

Yet many research questions remain open. For example, nobody knows exactly how many social bots populate social media, or what share of content can be attributed to bots ---estimates vary wildly and we might have observed only the tip of the iceberg. These are important questions for the research community to pursue, and initiatives such as DARPA's SMISC bot detection challenge, which took place in the Spring of 2015, can be effective catalysts of this emerging area of inquiry. 

Bot behaviors are already quite sophisticated: they can build realistic social networks and produce credible content with human-like temporal patterns. As we build better detection systems, we expect an arms race similar to that observed for spam in the past~\cite{heymann2007fighting}. The need for training instances is an intrinsic limitation of supervised learning in such a scenario; machine learning techniques such as active learning might help respond to newer threats. The race will be over only when the effectiveness of early detection will sufficiently increase the cost of deception. 

The future of social media ecosystems might already point in the direction of environments where machine-machine interaction is the norm, and humans navigate a world populated mostly by bots. 
We believe there is a need for bots and humans to be able to recognize each other, to avoid bizarre, or even dangerous, situations based on false assumptions of human interlocutors.\footnote{That Time 2 Bots Were Talking, and Bank of America Butted In --- \url{www.theatlantic.com/technology/archive/2014/07/that-time-2-bots-were-talking-and-bank-of-america-butted-in/374023/}}

\begin{acks}
The authors are grateful to Qiaozhu Mei, Zhe Zhao, Mohsen JafariAsbagh, and Prashant Shiralkar for helpful discussions.
\end{acks}

\newpage
\bibliographystyle{ACM-Reference-Format-Journals}
\bibliography{sigproc}

\received{June 201X}{201X}{201X}

\end{document}